%
% Interactions of strings and D-branes from M theory
% By Ofer Aharony, Jacob Sonnenschein and Shimon Yankielowicz
% February 1996 
%

\input harvmac.tex
\input epsf.tex

%\leftline{\bf{Preliminary Draft:\number\day/\number\month/\number\yearltd\ }}

% List of Journals
\def\np#1#2#3{Nucl. Phys. {\bf B#1} (#2) #3}
\def\pl#1#2#3{Phys. Lett. {\bf #1B} (#2) #3}
\def\prl#1#2#3{Phys. Rev. Lett. {\bf #1} (#2) #3}
\def\physrev#1#2#3{Phys. Rev. {\bf D#1} (#2) #3}

\def\mpl#1#2#3{Mod. Phys. Lett. {\bf A#1} (#2) #3}

% Reference list
\nref\towns{P. K. Townsend, ``The eleven dimensional supermembrane
revisited'', \pl{350}{1995}{184}, hep-th/9501068}%
\nref\witten{E. Witten, ``String theory dynamics in various dimensions'',
\np{443}{1995}{85}, hep-th/9503124}%
\nref\schwarz{J. H. Schwarz, ``An $SL(2,\bZ)$ multiplet of type II
superstrings'', \pl{360}{1995}{13}, hep-th/9508143; 
``Superstring dualities'', hep-th/9509148;
``The power of M theory'', \pl{367}{1996}{97}, hep-th/9510086; 
``M theory extensions of T duality'', hep-th/9601077}%
\nref\dfromm{P. K. Townsend, ``D-branes from M-branes'', hep-th/9512062}%
\nref\schmid{C. Schmidhuber, ``D-brane actions'', hep-th/9601003}%
\nref\first{E. Bergshoeff, E. Sezgin and P. K. Townsend, ``Properties of
the eleven dimensional supermembrane theory'', \pl{189}{1987}{75}}%
\nref\strominger{A. Strominger, ``Open $p$-branes'', hep-th/9512059}%
\nref\becker{K. Becker and M. Becker, ``Boundaries in M theory'',
hep-th/9602071}%
\nref\polchinski{J. Polchinski, ``Dirichlet branes and Ramond-Ramond
charges'', \prl{75}{1995}{4724}, hep-th/9510017}%
\nref\review{J. Polchinski, S. Chaudhuri and C. V. Johnson, ``Notes on
D-branes'', hep-th/9602052}%
\nref\al{F. Aldabe and A. L. Larsen, ``Supermembranes and superstrings with
extrinsic curvature'', hep-th/9602112}%
\nref\km{D. Kutasov and E. Martinec, ``New principles for string/membrane
unification'', hep-th/9602049}%
\nref\high{E. Bergshoeff, M. De Roo, M. B. Green, G. Papadopoulos and P. K.
Townsend, ``Duality of type II 7-branes and 8-branes'', hep-th/9601150}%
\nref\reducms{M. J. Duff, P. S. Howe, T. Inami and K. S. Stelle,
``Superstrings in $D=10$ from supermembranes in $D=11$'',
\pl{191}{1987}{70}}%
\nref\hankoh{S. K. Han and I. G. Koh, ``$N=4$ remaining supersymmetry
in Kaluza-Klein monopole background in $D=11$ supergravity theory'',
\physrev{31}{2503}{1985}}%
\nref\kapmic{D. M. Kaplan and J, Michelson, ``Zero modes for the
$D=11$ membrane and fivebrane'', \physrev{53}{3474}{1996},
hep-th/9510053}%
%\nref\dufflu{M. Duff and J. Lu, ``Black and super $p$-branes in diverse
%dimensions'', \np{416}{1994}{301}, hep-th/9306052}%
\nref\duff{M. J. Duff, S. Ferrara, R. R. Khuri and J. Rahmfeld,
``Supersymmetry and dual string solitons'', hep-th/9506057}%
\nref\kutasov{D. Kutasov, ``Orbifolds and solitons'', hep-th/9512145}%
\nref\hw{P. Horava and E. Witten, ``Heterotic and type I string
dynamics from eleven dimensions'', \np{460}{1996}{506}, hep-th/9510209}%
\nref\orbifolds{K. Dasgupta and S.
Mukhi, ``Orbifolds of M theory'', hep-th/9512196; E. Witten, ``Five-branes
and M theory on an orbifold'', hep-th/9512219; A. Sen, ``M theory on $(K3
\times S^1) / \bZ_2$'', hep-th/9602010; A. Kumar and K. Ray, ``M-theory on
orientifolds of $K3 \times S^1$'', hep-th/9602144}%
\nref\dlp{J. Dai, R. Leigh and J. Polchinski, ``New connections between
string theories'', \mpl{4}{1989}{2073}}%
\nref\aspinwall{P. S. Aspinwall, ``Some relationships between dualities in
string theory'', hep-th/9508154}%
\nref\bound{E. Witten, ``Bound states of strings and $p$-branes'',
\np{460}{1996}{335}, hep-th/9510135}%
\nref\tseytlin{A. A. Tseytlin, ``Self duality of the Born-Infeld action and
Dirichlet 3-branes in type IIB string theory'', hep-th/9602064}%
\nref\gg{M. B. Green and M. Gutperle, ``Comments on three-branes'',
hep-th/9602077}%
\nref\verlinde{E. Verlinde, ``Global aspects of electric-magnetic
duality'', \np{455}{1995}{211}, hep-th/9506011}%
\nref\scatter{I. R. Klebanov and L. Thorlacius, ``The size of $p$-branes'',
hep-th/9510200; S. S. Gubser, A. Hashimoto, I. R. Klebanov and J. M.
Maldacena, ``Gravitational lensing by $p$-branes'', hep-th/9601057}%
\nref\comments{E. Witten, ``Some comments on string dynamics'', to appear
in the proceedings of Strings '95, hep-th/9507121}%
\nref\tachyon{M. B. Green, ``Pointlike states for type IIB superstrings'',
\pl{329}{1994}{435}, hep-th/9403040;
T. Banks and L. Susskind, ``Brane-antibrane forces'', hep-th/9511194}%
\nref\pw{J. Polchinski and E. Witten, ``Evidence for heterotic-type I
string duality'', \np{460}{1996}{525}, hep-th/9510169}%
\nref\vafa{C. Vafa, ``Evidence for F-theory'', hep-th/9602022; D. R.
Morrison and C. Vafa, ``Compactifications of F-theory on Calabi-Yau
threefolds - I'', hep-th/9602114}%

% Useful definitions
\def\bZ{{\bf Z}}

% Title page

\Title{hep-th/9603009, TAUP-2324-96}
{\vbox{\centerline{Interactions of strings and D-branes from M
theory}}}

\bigskip

\centerline{\bf Ofer Aharony\foot{Work supported in part by the 
US-Israel Binational Science Foundation, by GIF -- the German-Israeli
Foundation for Scientific Research, and by the Israel Academy of 
Science.}$^,$\foot{Work supported in part by the Clore Scholars Programme.
Address after Sept. 1, 1996 : Department of Physics and Astronomy, Rutgers
University, Piscataway, NJ 08855-0849. 
E-mail address : oferah@post.tau.ac.il.}, Jacob Sonnenschein$^1$
and Shimon Yankielowicz$^1$}
\vglue .5cm
\centerline{School of Physics and Astronomy}
\centerline{Beverly and Raymond Sackler Faculty of Exact Sciences}
\centerline{Tel--Aviv University}
\centerline{Ramat--Aviv, Tel--Aviv 69978, Israel}

\bigskip\bigskip

\noindent
We discuss the relation between M theory and type II string theories. We
show that, assuming ``natural'' 
interactions between membranes and fivebranes in
M theory, the known interactions between strings and D-branes in type II
string theories arise in appropriate limits. Our discussion of the
interactions is purely at the classical level. 
We remark on issues associated with the M theory approach to enhanced 
gauge symmetries, which deserve further investigation.

\Date{3/96}

% Paper
\newsec{Introduction}

The idea that string theory should be reformulated in terms of an eleven 
dimensional theory, whose low energy limit is eleven dimensional
supergravity, has been revived in the past year in the form of M
theory. Even though the fundamental formulation of this theory is still not
known, the assumption of its existence
has led to a simpler understanding of many types of string
dualities whose origin is otherwise obscure. It is thus tempting to believe
that a consistent quantum description of M theory does indeed exist. This
is also suggested by the possible ``definition'' of M theory as the strong
coupling limit of the type IIA string theory.

In this paper we discuss the relationship between M theory and
type II string theories. We will try to learn from string theory
more about the fundamental formulation of M theory. The original motivation
for an eleven dimensional origin for string theory came from the existence
of a supergravity theory in eleven dimensions, which reduces to the type
IIA supergravity theory upon dimensional reduction to ten dimensions. This
suggests that the low energy effective description of M theory should be
eleven dimensional supergravity, but says nothing about its fundamental
objects. The major advance in the past year came from the realization that
the BPS saturated $p$-branes of string theory, both perturbative and
solitonic, may all be described by assuming the existence of a membrane and
a fivebrane in M theory \refs{\towns,\witten,\schwarz}. 
The membrane couples to the 3-form field of eleven
dimensional supergravity, while the fivebrane couples to its dual. The
low-energy effective actions for the extended BPS saturated $p$-branes in
string theory were also connected to M theory \refs{\dfromm,\schmid}, 
by assuming that the
action of the membrane is just the supermembrane action \first, and that
the fivebrane is the solitonic fivebrane of eleven dimensional
supergravity. All of these relations were derived 
at the level of the effective
action. For this it does not matter whether the membrane, for instance, is
fundamental (as in supermembrane theory), or whether it is a solitonic
$p$-brane of some deeper fundamental theory. The first possibility is
supported by the fact that the solitonic solution of eleven dimensional
supergravity corresponding to the membrane is singular.

Our goal in this paper is to take the relation between M theory and
string theories beyond the level of the spectrum, and discuss the
interactions in the two theories. We will assume that the interactions of
membranes in M theory are correctly described by supermembrane theory, and
that the fivebranes in M theory may be described as D-branes on which
membranes may have boundaries \refs{\strominger,\dfromm,\becker}. Our
discussion of RR $p$-branes in string theory will be based on their
formulation by Polchinski \polchinski\  as boundaries for open strings (see
\review\ for a recent review).
Of course, since we have no quantum
formulation of supermembrane theory, our discussion throughout this paper
will be purely classical. Since interactions necessarily involve
non-BPS saturated states, quantum corrections could be important.
However, at the classical level we will be able
to show, using known results, 
that by assuming the above natural 
interactions in M theory we get the correct
interactions between strings and $p$-branes (for $p \leq 6$) when we go
over to the string theory limit. 
We believe that this presents stronger evidence for
the existence of a quantum M theory, whose classical limit gives
supermembrane theory, at least as an effective action. This could be some
version of supermembrane theory which avoids the problem of the continuous
spectrum (such as a membrane with thickness \refs{\towns,\al}), a string
theory for which the membrane action is a low-energy effective description
\km, or a
completely different theory which we have not yet been able to imagine.

In section 2 we describe the known relations between M theory and type IIA
string theory, and show how they lead to the same interactions in the two
theories. In section 3 we perform the same analysis for the type IIB string
theory in ten dimensions. 
In section 4 we discuss some issues, associated with the description
of D-branes in string theory, whose M theory interpretation is still
unclear. Understanding these issues may help toward the formulation of M
theory. We end in section 5 with a summary and some open questions.

\newsec{Interactions of type IIA string theory from M theory}

The simplest relation between M theory and a string theory arises when we
compactify one dimension, which we will denote by $X^{11}$, on a circle of
radius $R_{11}$. In the limit of small $R_{11}$, this is believed to lead
to the ten-dimensional type IIA string theory, with a string coupling
$e^{\phi_A}$ proportional to $R_{11}^{3/2}$ \refs{\towns,\witten}
(we will ignore numerical
constants throughout this paper, as well as the tensions of the string and
of the membrane which set the scale in relations of this type). 
We will begin this section
by reviewing the evidence for this relationship, and continue by
generalizing it also to the interactions of the type IIA string with
D-branes.

The simplest evidence for the relation between M theory and type IIA
strings comes from the low-energy effective action. It is just the
statement that the dimensional reduction of 11 dimensional supergravity
gives 10 dimensional type IIA supergravity, when we identify the 11
dimensional metric with the 10 dimensional metric, RR 1-form and dilaton by
\eqn\metrics{g^{(11)}_{mn} dx^m dx^n = e^{-2\phi_A / 3} g^{(10)}_{\mu \nu} 
dx^{\mu} dx^{\nu} + e^{4\phi_A / 3} (dx^{11} - A_{\mu} dx^{\mu})^2}
(throughout this paper we will use $m,n,\cdots$ for 11 dimensional
indices, $\mu,\nu,\cdots$ for 10 dimensional indices, and $\alpha,
\beta,\cdots$ for 3 dimensional indices on the worldvolume of the
membrane). We assume, as usual,
that none of the background fields depend on $X^{11}$. The 
2-form $B_{\mu \nu}$ and
3-form $A_{\mu \nu \lambda}$ of type IIA string theory
are related to the 3-form $A^M_{mnk}$
of 11 dimensional supergravity simply
by $A_{\mu \nu \lambda} = A^M_{\mu \nu \lambda}$ and $B_{\mu \nu} = 
A^M_{\mu \nu (11)}$. For this relation between the theories
we need only to use the fact that the low-energy 
effective description of M theory is given by 11 dimensional
supergravity.

Additional evidence for the relation between M theory and type IIA string
theory comes from the identification \refs{\towns,\witten,\schwarz} 
of the BPS saturated $p$-branes of the two
theories, for $p \leq 6$\foot{The issue of $p$-branes with $p > 6$ and
their relation to M theory is more complicated \refs{\schwarz,\high} and
will not be discussed here.}, and of the effective actions describing
these $p$-branes. For this we need to assume a particular spectrum of
$p$-branes in the 11 dimensional M theory, as well as the effective actions
describing them. It turns out that the identification goes through if
we use the membrane and fivebrane\foot{To avoid unnecessary confusion, we
will use the word ``fivebrane'' to describe the 5-brane of M
theory, and the word ``5-brane'' to describe the 5-branes of string
theory. The word ``membrane'' or ``supermembrane''
will denote the 2-brane of M theory, and the
word ``2-brane'' will denote the 2-brane of string theory.} of supermembrane 
theory \first\ as the only BPS
saturated $p$-branes of M theory. 

First, it has long been known \reducms\  that the double dimensional
reduction of the supermembrane worldvolume theory gives the type IIA 
superstring worldsheet theory, at least on the classical level (our
discussion throughout this paper will be purely classical, since we do not
know how to quantize supermembrane theory). Thus, a
supermembrane for which one worldvolume coordinate is always wrapped around
$X^{11}$ is, at least classically and in the $R_{11} \to 0$ limit, 
exactly the same as a type IIA superstring. 

A more complicated identification relates a supermembrane which is not
wrapped around $X^{11}$ with the D-2-brane of type IIA string theory. The
two effective
actions in this case turn out to be the same \refs{\dfromm,\schmid}
if we perform a three
dimensional duality transformation on the worldvolume of the membrane,
transforming the scalar $X^{11}$ into a vector field $A_{\alpha}$. Since
the only dependence of the supermembrane action on $X^{11}$ is through its
kinetic term, we can treat $w_{\alpha} = \del_{\alpha} X^{11}$ as a
fundamental field if we add to the action a Lagrange multiplier term of 
the form 
$\epsilon^{\alpha \beta \gamma} \Lambda_{\alpha} \del_{\beta} w_{\gamma}$.
By solving the equation of motion of $w_{\gamma}$ we find that $F_{\alpha
\beta} = \del_{\alpha} \Lambda_{\beta} - \del_{\beta} \Lambda_{\alpha}$
behaves as a gauge field strength (together with $B_{\alpha \beta}$ defined
below), and the action of the supermembrane becomes
exactly the effective action of the Dirichlet 2-brane. This includes the
Born-Infeld term $\sqrt{-\det(G_{\alpha \beta} + F_{\alpha \beta} -
B_{\alpha \beta})}$, where 
$G_{\alpha \beta} = \del_{\alpha} x^{\mu}
\del_{\beta} x^{\nu} g^{(10)}_{\mu \nu}$ and 
$B_{\alpha \beta} = \del_{\alpha} x^{\mu}
\del_{\beta} x^{\nu} B_{\mu \nu}$. The equation of motion of $w_{\alpha}$
sets
\eqn\walpha{w_{\alpha} \sim e^{-\phi_A}
\epsilon_{\alpha \beta \gamma} (F^{\beta \gamma} - B^{\beta \gamma})}
up to factors involving the metric on the supermembrane which we ignore.
The fermionic terms in \walpha, and throughout the paper, are suppressed,
but we do not expect their analysis to present any fundamental difficulties.
They should just supersymmetrize the bosonic terms.
Note that the holonomy of the gauge field, when the 2-brane has
topologically non-trivial cycles, is not determined by this
transformation. Other
global issues involved in this transformation are discussed below.

The D-4-brane and NS 5-brane of type IIA string theory are similarly
related to the wrapped or unwrapped ``magnetic'' fivebranes of M 
theory \refs{\schwarz,\dfromm}. 
In this case an exact expression for the effective
worldvolume action of the
fivebrane is not available, since it involves an anti-self-dual 3-form 
field strength.
However, the fields along the D-4-brane and the NS 5-brane are correctly
given by the double and single dimensional reductions of the ``magnetic''
fivebrane fields. The
tensions of these $p$-branes, as well as the tensions of the string and
membrane described in the previous paragraphs, also have the correct
dependence on the string coupling constant \schwarz, using the relations
between their descriptions in M theory and in string theory
and equation \metrics.

The other BPS saturated $p$-branes of the type IIA string theory with $p
\leq 6$ are the
Dirichlet 0-brane and 6-brane. These arise in M theory as ``electric'' and
``magnetic'' Kaluza-Klein excitations \refs{\towns,\witten,\schwarz}. 
The 0-branes
are characterized just by their mass, which is the same in both of their
interpretations \witten. For the 6-branes the tensions in both
interpretations are proportional (in the string metric)
to $e^{-\phi_A}$, and it is known that the Kaluza-Klein monopole in 11
dimensional supergravity breaks half of the supersymmetries \hankoh.
It should be possible to derive the low energy effective action of the
6-brane from the Kaluza-Klein monopole solution to the supergravity
theory, as done for the membrane and the fivebrane in
\kapmic. Obviously there are 3 zero modes corresponding to transverse
motion of the 6-brane, and the other 5 bosonic zero modes should
correspond to the Born-Infeld vector field in the worldvolume 
of the 6-brane. We
expect this vector field to arise from fluctuations in the 3-form
field $A^M_{mnk}$ around the monopole solution, 
in the same way as the 2-form arises in the worldvolume theory
of the fivebrane \kapmic, but this has not yet
been verified as far as we know.

The next step after identifying the spectrum of the two theories should
involve checking the interactions of the various $p$-branes. The basic
interactions of supermembranes in supermembrane theory are geometrical
splittings and joinings of the membranes, like in superstring theory. An
important difference between the two theories is that in 11 dimensional
supermembrane theory there is no dimensionless coupling constant, so any
``interaction vertex'' is necessarily of order one, and it is not clear how
to define a perturbative expansion. However, once we introduce a small
compactification radius, we have a small dimensionless coupling in the
theory (which is just the radius in units determined by the supermembrane 
tension), and it is meaningful to talk about vertices of $p$-branes 
which wrap
around this small dimension, since these become weakly coupled.  Another
difference between membranes and strings is that free membranes can
smoothly change their topology (in space). 
Such topology changes will be involved in
some of the interactions described below, such as the generation of a
``virtual'' open string on a 2-brane.

Let us start with the string and 2-brane of type IIA string theory, both of
which arise from the supermembrane in 11 dimensions. A diagram
for the interaction of strings (in 10 dimensions)
may be transformed into a diagram of
membranes in M theory just by replacing each point in the diagram by a
circle wrapped around the eleventh dimension. Since the tensions of the
string and membrane are related simply by $T_S = T_M 2 \pi R_{11}$ 
(when both are written in the same metric \schwarz), it is obvious from the
Nambu-Goto form of the action that the action associated with any diagram
is the same in both descriptions. The only subtle issue here is the
dependence of the diagram on the string coupling (or the dilaton), since
when one performs the double dimensional reduction of the supermembrane
action \reducms, 
one gets the string action coupled to the $g^{(10)}_{\mu \nu}$ and $B_{\mu
\nu}$ fields but not to the dilaton. This is, of course, to be expected
since this reduction is purely classical, and the dilaton term is of higher
degree in $\alpha'$ than the others. From the point of view of M theory the
dilaton is just a Kaluza-Klein scalar field.
We expect a correct quantum treatment
of this reduction to give also the dilaton term.
It is, however, not clear how to do this
rigorously, since the supermembrane worldvolume action is generally not
renormalizable.

The other interaction of membranes which we can discuss at weak coupling is
the interaction of a string with a 2-brane in the type IIA string theory.
In the string theory, this interaction is described \polchinski\ simply by
allowing the string to end on a 2-brane, as in figure 1, where we took the
2-brane to be spherical.

\epsfbox{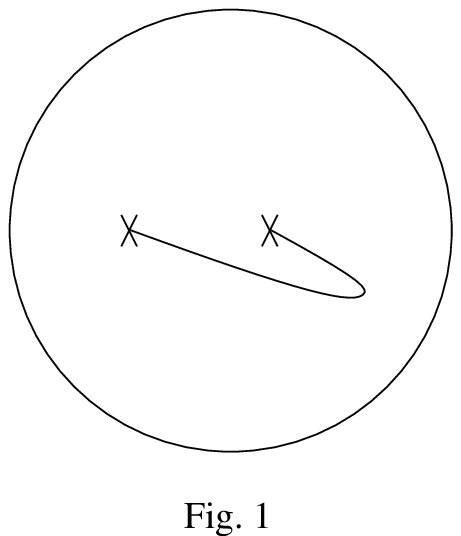}

The end of the string
behaves (inside the 2-brane) as a particle
charged with respect to 
the gauge field on the surface of the 2-brane. In order to get
the corresponding interactions in M theory,
we should look at diagrams of membrane interactions which include a
region in which the membrane is wrapped around $X^{11}$. The string in
figure 1 should be replaced by such a wrapped membrane, joined smoothly to
the rest of the membrane.
When we take $R_{11}$ to zero, this becomes just the
interaction of the string with the 2-brane. The only thing left to check is
that the end of the ``string'' behaves like a charged particle with respect
to the gauge field living in the unwrapped part of the membrane. The end of
the ``stringy'' part of the membrane is topologically
a circle in the rest of the
membrane, which is wrapped around $X^{11}$. Thus, if we perform an integral
$\oint dx^{\alpha} \del_{\alpha} X^{11}$ around the end-point of the
string, it will be a non-zero constant (namely, $2\pi R_{11}$). However, the
identification of the 2-brane action with the membrane action \walpha\
means that this is proportional to $\oint \epsilon_{\alpha \beta \gamma}
(F^{\beta \gamma} - B^{\beta \gamma}) dx^{\alpha}$ which is just the electric
flux emanating from the end-point of the string. Thus, we have shown that the
interactions of strings and 2-branes in type IIA string theory are
correctly described by M theory. In both cases the
effective action describing the interaction includes a part
corresponding to the separate actions of the string and the membrane (which
are identified as above), and a contribution from the boundary of the
string, or the interaction region between the membranes, which (in the
$R_{11} \to 0$ limit) is just the
action for the motion of a charged particle.

Next, we should add the fivebranes of M theory. These play the 
same role in M
theory as the D-branes in string theory \refs{\strominger,\dfromm,\becker}.
The membranes can have a boundary along a fivebrane, and it behaves (in the
$5+1$ worldvolume theory of
the fivebrane) as a string which is charged with respect to 
the 2-form field living
on the fivebrane (this is a limit of the dyonic string of \duff). 
 From here we can easily see the interactions of the type
IIA string with 4-branes and 5-branes. The interaction of the type IIA
string with a 4-brane is just the double dimensional reduction of the
interaction of a membrane with a fivebrane \dfromm, 
with both worldvolumes wrapped
around $X^{11}$. The boundary of the membrane becomes the boundary of a
string, and it is charged with respect to 
the gauge field of the 4-brane (which is the
dimensional reduction of the 2-form field on the fivebrane), as desired.
We find no similar interactions between the
string and the 5-brane, since the end of a ``stringy'' membrane is a circle
wrapped around $X^{11}$, and there are no such cycles in the unwrapped
fivebrane. This is consistent with the absence of such interactions between
the string and the NS 5-brane in the type IIA string theory. There will be
more complicated interactions between the string and the 5-brane when the
radius of the eleventh dimension is non-zero, which will lead to scattering
of strings and 5-branes, and it would be interesting to compare these with
the corresponding interactions in string theory. To compute these we
require a conformal field theory description of the NS 5-brane, perhaps
along the lines of \kutasov.

The other interactions which have a simple description in the type IIA
string theory are between the string and the 0-brane and 6-brane. The
0-brane can be viewed simply as a collapsed membrane \refs{\dfromm,\becker},
ensuring that its interactions are also the same in string theory and M
theory. In fact, since an open string with both ends on a 0-brane is just a
closed string, there are actually no purely bosonic
interactions between strings and
0-branes, at least at the level of the low-energy effective action.
This is clear both in the string picture, since there are no new string
sectors, and in the membrane picture, since the 0-brane may be viewed as
a membrane whose size has gone to zero.
The Kaluza-Klein ``magnetic'' 6-brane of M theory on a circle may be
viewed as a non-trivial embedding of the $S^1$ of the eleventh
dimension in the other dimensions. In this embedding, the radius of
the eleventh dimension goes to zero at the position of the 6-brane
(this is also clear from the type IIA picture, where the dilaton $e^{\phi}$
vanishes at the position of the 6-brane). Thus, a membrane wrapped
around the eleventh dimension may end on the 6-brane, and in the
string theory limit this would give a string ending on a Dirichlet
6-brane. An analysis of the effective field theory on the 6-brane
should lead to the endpoint of this string being charged with respect
to the electric field on the 6-brane. In the 11 dimensional
supergravity this electric field arises from zero modes of the monopole
solution corresponding to turning on the 3-form field, under which the
membrane is charged. Thus, this seems reasonable, but we have not
checked it rigorously.
%We do not yet know how to see the electric field on the 6-brane
%from M theory, so we cannot check that the endpoint of the string is
%electrically charged under it, as it should be.
%We did not check the interactions of the Kaluza-Klein ``magnetic''
%6-brane of M theory on a circle with membranes, but we expect that they
%will also give rise to the correct interactions of strings with 6-branes.
Other interactions (such as interactions of 2-branes with themselves)
cannot be seen at weak coupling in string theory, hence we cannot compare
them. We expect that M theory will also have interactions between fivebranes,
but we do not know how to describe these in M theory or in string theory,
so obviously we cannot compare them.

\newsec{Interactions of type IIB string theory from M theory}

The other string theory which is simply related to M theory is the type IIB
theory (for other theories we need to put M theory on an orbifold, and it
is not clear how the twisted sectors should be defined in M theory, though
there is by now a certain amount of experience with orbifolds
\refs{\hw,\orbifolds} 
which should hint at the appropriate definition). To reach type IIB string
theory, we compactify two of the dimensions of M theory on a torus. By the
analysis of the previous section, this is just type IIA string theory on a 
circle,
which is known to be related by T duality to type IIB string 
theory on a circle \dlp. By performing this duality, we find \schwarz\ 
that M theory on a torus of area $A_T$
and of modular parameter $\tau$ seems equivalent to type IIB string 
theory on a
circle of radius $R_B \sim A_T^{-3/4}$, with a complex coupling constant
$\lambda = \chi + i e^{-\phi_B}$ (where $\chi$ is the RR 0-form or axion)
which equals $\tau$. Note that here we do not need to take the compact
dimensions to zero in order to get a weakly coupled string theory, since
the string coupling depends only on the modular parameter of the torus. We
should, however, take the size of the torus to zero (with a constant
modular parameter) in order to get the type 
IIB string theory in ten dimensions
(since then $R_B \to \infty$), and this is the case we will analyze here. 
This means that in the context of M theory,
the ``stringy'' description is in fact
exact for type IIB theory in ten dimensions, while for type IIA it is only
a small coupling (small radius) approximation.
The $SL(2,\bZ)$ duality of the type IIB
string theory obtains through this construction a geometrical
interpretation, as the group of modular transformations of the torus on
which M theory was compactified \refs{\aspinwall,\schwarz}.

The identification between M theory on a torus and type IIB theory is based
on the same type of evidence as described above for the type IIA theory. In
fact, since the relation of type IIA and type IIB theories on a circle via
T duality is well understood (also for the $p$-branes of the two theories),
the identification of the type IIA theory is apparently sufficient in order
to identify the type IIB theory as well. However, the weak coupling limits
of the type IIB theory (in which we understand it as a string theory) are not
necessarily related to weak coupling limits of the type IIA theory. Thus, more
checks are actually necessary, since our identification of the interactions
above was just for weak coupling. In any case, some interactions 
look different
in the type IIB theory, so it is worthwhile to repeat the whole analysis
above on the description of the $p$-branes and of their interactions.

The simplest $p$-branes in the type IIB theory are the bound states of 
strings
and D-strings \bound. For every pair of coprime integers $(n,m)$ there
exists a BPS saturated bound state of $n$ fundamental strings and $m$ 
D-strings, whose worldsheet low-energy effective action is essentially 
that of the
type IIB string. In the M theory, these are \schwarz\ identified with
membranes with the topology of a torus, one of whose cycles is
wrapped $n$ times around one cycle of the space-time torus and 
$m$ times around the other cycle\foot{When $n$ and $m$ are not coprime,
these states are only marginally stable \schwarz\ and are not expected to
exist as bound states.}.
As described in the previous section, this gives a string with the type 
IIA string action in 9 dimensions, which after the T duality becomes the
type IIB string action as desired. Membranes which are not wrapped around
any cycle of the torus become wrapped around the circle of type IIB theory
after the T duality transformation, so that they are not visible in ten
dimensional type IIB theory (we will limit ourselves here only to a
discussion of the states which remain in the limit of $R_B \to \infty$).
Membranes which are completely wrapped around the torus are identified
\schwarz\ with Kaluza-Klein particles of type IIB theory on a circle. In
the ten-dimensional limit these become part of the ten dimensional fields.
Thus, the $(n,m)$ strings are all that remain
of the membrane after compactifying the space-time on a very small torus. 
The fivebrane, on the
other hand, must be wrapped around both cycles of the torus to give a state
of ten dimensional type IIB theory (it must of course have an appropriate
topology for this). The difference between the two arises from the
different properties of NS and R branes under T duality, as can be seen
from their type IIA description. The fivebrane wrapped around the torus
gives an $SL(2,Z)$ invariant 3-brane in the type IIB theory, which should be
identified with the D-3-brane of this theory. The effective
action of the D-3-brane
is the supersymmetric generalization of the Born-Infeld action in 3+1
dimensions. Its low-energy limit is just $N=4$ $U(1)$ gauge theory in 3+1
dimensions \refs{\tseytlin,\gg}, which is known to be $SL(2,\bZ)$ invariant.
As far as we know,
this action has not yet been derived from the double dimensional
reduction on a torus 
of the fivebrane of M theory, but essentially this should be
just the supersymmetric generalization of the dimensional reduction of
\verlinde. Using this reduction, the integrals of the 2-form field of the 
fivebrane around the two cycles of the
torus give the electric and magnetic fields on the worldvolume of the
3-brane (recall that the field strength
of the 2-form is anti-self-dual, so that the two integrals do
not give independent 1-forms). If the fivebrane is wrapped $n$ times around
the torus, the low-energy effective description should 
be given by the $N=4$ $U(n)$ gauge theory \refs{\bound,\gg}, which is also
believed to be $SL(2,\bZ)$ invariant. 

The other BPS saturated $p$-branes in the type IIB theory (with $p \leq 6$)
are the 5-branes, which are also labeled by two coprime integers $n$ and
$m$. These arise in M theory as Kaluza-Klein ``magnetic'' branes \schwarz, 
like the 6-brane in the type IIA theory.

Next, let us move on to the interactions, beginning with the interactions
between a string and a 3-brane in the type IIB theory. These arise from
membrane-fivebrane interactions in M theory, like the string-4-brane
interactions described in the previous section. 
In this case any $(n,m)$ string can end on
the 3-brane (which is a fivebrane wrapped around the torus), 
since the end-cycle of the appropriate ``stringy'' membrane 
is wrapped
around the $(n,m)$ cycle of the torus, and all such cycles exist in the
fivebrane when it is wrapped around the torus. The end-point of such a
string describes (in the 4 dimensional effective theory of the
3-brane) a dyon with electric charge $n$ and magnetic charge $m$ \gg. This
arises from the coupling of the boundary of the membrane to the 2-form of the
fivebrane, as in the previous section.

The interactions of strings with 5-branes are exactly analogous to
the interactions of strings with 6-branes described in the previous
section. At the Kaluza-Klein 5-brane, the appropriate $(n,m)$ cycle of
the torus goes to zero size (in the 11 dimensional metric), so that a
membrane wrapped around this cycle may end on the 5-brane. In the
string theory limit this gives an $(n,m)$ string ending on an $(n,m)$
5-brane.

The interactions of strings of a particular type $(n,m)$, when the coupling
is such that these strings are weakly coupled, arise from M
theory in the same way as the interactions of type IIA strings
described above. 
More interesting is the interaction between strings of
different types, for instance between a $(1,0)$ fundamental string and a
$(0,1)$ D-string when the fundamental string is weakly coupled (the
generalization to general strings is straightforward). Let us look at a
generic interaction in the type IIB string theory, such as the scattering
of a string on a D-string \scatter, described (in spacetime) in figure 2.

\epsfbox{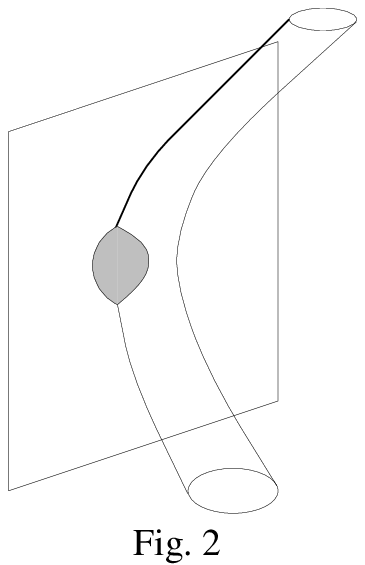}
 
When a fundamental string has end-points on a D-string, these are charges
on the worldsheet of the D-string, so that there is a constant electric
field between the two end-points. Thus, according to the description of
D-strings in \bound, the portion of the D-string between the two end-points
is actually a $(1,-1)$ string, and the time slices of the interaction
actually look as in figure 3a (taking the D-string also to be a finite
closed string). How can we understand such an interaction
from M theory ? In the beginning we have two toroidal membranes, one with a
cycle wrapped around the $(1,0)$ cycle of the spacetime torus and another
with a cycle wrapped around the $(0,1)$ cycle of the spacetime torus. Then,
they join together to give a genus 2 surface, as in figure 3b. 

\epsfbox{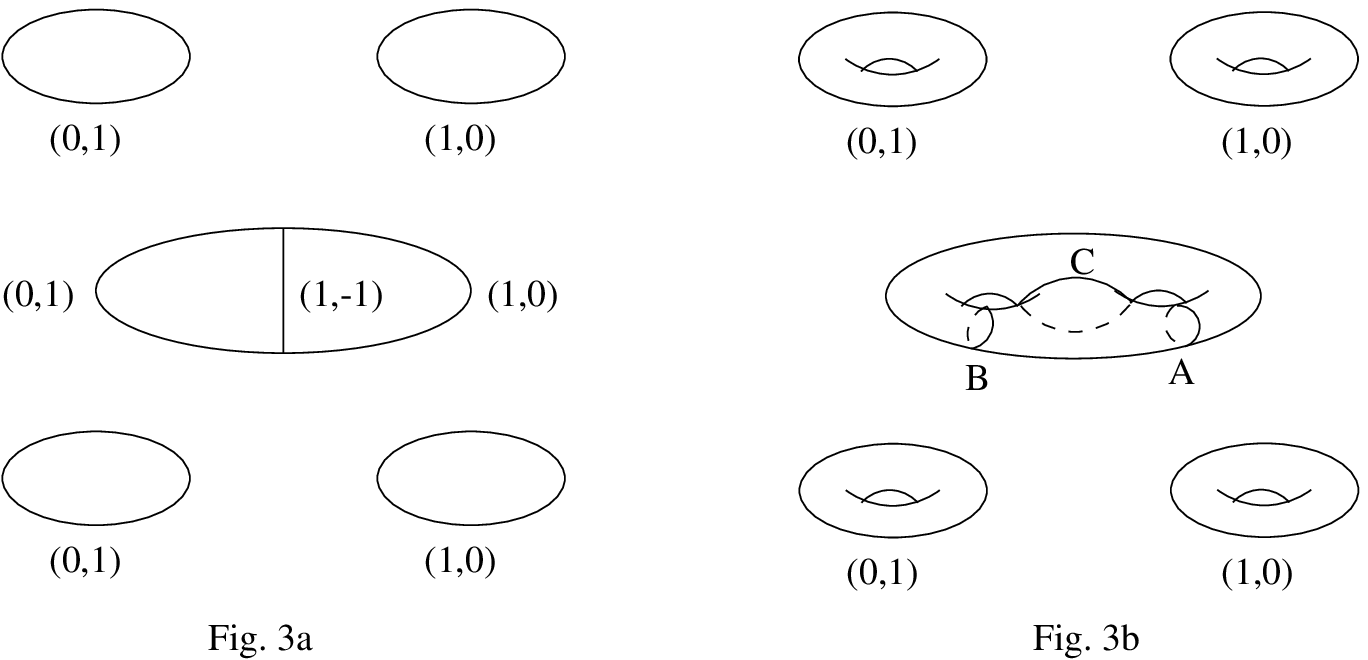}

When we take
the spacetime torus to zero, the stringy description of this
interaction becomes exactly that of figure 3a. The ``middle'' cycle
(denoted by C in figure 3b) of
the genus 2 surface is homologically the difference between the cycles A
and B, and thus is wrapped around the $(1,-1)$ cycle of the spacetime torus,
as desired. Thus, the basic interactions of strings and D-strings are
also correctly described by the basic interaction between membranes in M
theory. 

\newsec{Remarks on enhanced gauge symmetries in M theory}

The description of D-strings in M theory, as in the previous section,
seems very different from their
description in \bound. Witten described the effective theory of the $(n,m)$
string as a $U(n)$ gauge theory with $m$ quarks at infinity. However, since
for coprime $n$ and $m$ this theory is believed to have a mass gap, the 
low-energy 
degrees of freedom (which are the only ones we are discussing here)
are only those in the $U(1)$ gauge superfield (including the scalars
corresponding to the position of the D-string), which we get also from the
M theory description. As we discussed in the previous section, we can
replace one of the compact dimensions of the torus by a gauge field on the
worldvolume of the membrane. When we go over to the string picture, this
becomes a gauge field on the worldsheet of the D-string, which has (in the
``stringy'' limit) a
constant value corresponding to the winding number around the appropriate
compact dimension. Thus, we get exactly the same description as in
\bound.

However, we should also be able to discuss
configurations of $n$ $(1,0)$ D-strings (for instance). According to
\bound, such configurations are described by a $U(n)$ gauge theory on the
worldsheet of the D-string. How does this arise in M theory ? In string
theory, when we have two parallel D-strings, we get a
W-boson state from an open fundamental string connecting the two. In M
theory, therefore, we expect a ``wormhole'' configuration between 
the two membranes to give rise to the enhanced gauge
symmetry. Whenever two membranes touch each other, the
worldvolume scalars describing the position of the two membranes should 
become a
$2\times 2$ matrix \bound. Perhaps this can be understood by examining
the possible interactions of the two membrane sheets
at this point. Even though
in string theory there is no geometrical
interpretation of this phenomenon,
we believe that such an interpretation may
be possible in M theory, but we have not been able to find it.
Another possibility is that the enhanced gauge symmetry may
only be understood in M theory at the quantum level.
This is reasonable, since the additional states appear in the
effective action only when the distance between the D-branes is much
smaller than the appropriate ``Planck length''.
Since we do not know how to quantize the M theory, it is difficult to
explore this possibility.
Note that the same mechanism should give the enhanced gauge symmetry for
the D-strings in the type IIB theory and for the D-2-branes of the type IIA
theory discussed in the previous section. 
For fivebranes in M theory the enhanced gauge symmetry arises \strominger\
from tensionless strings \comments\ 
which appear when the distance between the fivebranes goes to zero.
The winding states of these strings become additional gauge
bosons. In string theory this leads to the enhanced gauge
symmetries which occur for the D-4-branes of the type IIA theory and for the
D-3-branes of the type IIB theory. Perhaps we may be able to understand
the analogous problem for membranes in a similar way,
since the boundary of one membrane on another membrane is
also a string, whose tension goes to zero when two membranes approach each
other.

The enhanced symmetry described above should arise only for two membranes
which have the same orientation. Membranes with different orientations go
over, in the string theory, to a D-brane anti-D-brane pair, and then we
expect a tachyonic instability to develop when they approach each other
\tachyon. 
Obviously, the geometry of the connecting ``wormholes'' is different in
the two cases. Hence, it is possible that such an instability may indeed
exist for oppositely oriented membranes, 
but we do not know how to derive it from M theory. Perhaps
it can only be seen in the quantum theory which we do
not yet know how to formulate.

At this point we should comment that if the $SL(2,\bZ)$ symmetry of type
IIB string theory is exact
(this is obvious if the type IIB string is indeed exactly described by M
theory on a torus), there should be no difference between the
fundamental string and the D-string. Any apparent difference between the
two must come from the fact that one of them is (in the usual description)
weakly coupled, while the other is strongly coupled. Differences which do
not depend on the string coupling, for instance in
the spectrum of BPS saturated
states, cannot exist. At weak coupling it seems that the string world-sheet
action is fundamental, while the D-string worldsheet action arises from
open strings coupled to the D-string. A reciprocal picture should give
the correct description at strong coupling. 
As described in the previous section, the
interactions of the strings and of the D-strings seem to be the same.
Thus, the gauge field which appears
in the effective theory of the D-string and not in the fundamental theory
of the string does not play any dynamical role. 
One of the apparent differences
between strings and D-strings is that for $n$ D-strings which sit on top of
each other we get an enhanced $U(n)$ gauge group, while no similar
phenomenon is known for fundamental strings. This leads, for instance, to
the fact \review\ that there is no distinction between a D-string wrapped
twice around a space-time circle or two D-strings wrapped once around the
same circle. For fundamental strings the corresponding perturbative states
are clearly distinct.
So far we have not understood how to get this enhanced gauge
group from M theory, but it seems that if it arises for D-strings it should
arise also for fundamental strings, since the description of the two in M
theory should be the same. It does not seem
reasonable, though we have not been able to rigorously
rule out this possibility,
that this gauge symmetry would only arise for strongly coupled strings.
Of course, in strong coupling the additional states giving rise to the
enhanced gauge symmetry may be described by open D-strings connecting the
fundamental strings. This is just the $SL(2,\bZ)$ dual of the picture
described in \bound.

\newsec{Summary and future directions}

In this paper we examined the correspondence between the interactions in M
theory and in type II string theories. We found that the basic interactions
between membranes and membranes and between membranes and fivebranes in M
theory give rise to the known interactions of strings and D-$p$-branes
(with $p \leq 6$) in type II string theory, in appropriate weak coupling
limits. The understanding of $p$-branes with $p > 6$ in M theory is
still not clear.
%Our discussion was limited to the $p$-branes which arise from the
%membrane and fivebrane in M theory. It would be interesting to generalize
%the discussion to include also higher $p$-branes, such as those which arise
%as Kaluza-Klein states. 
In ten dimensions $p$-branes with $p > 6$ affect
the asymptotic form of the metric (at least when they are flat), 
so that they cannot appear in flat ten
dimensional spacetime. Thus, to discuss them we need to
study M theory on curved space. 
Another possible generalization of our work is to
orbifolds of M theory \refs{\hw,\orbifolds}, 
which are believed to describe heterotic
strings. The analysis of the bulk interactions in these theories is similar
to the analysis we performed here, but new issues arise at the
orbifold fixed points, which are beyond the scope of this paper.
Type I strings also appear in some of 
these theories, which are not BPS saturated
states \refs{\pw,\hw}. 
It is not clear how to analyze the interactions of these strings
in M theory.

Our analysis was limited, by the level of our knowledge of M theory, 
to a classical analysis of low-energy effective actions. Such an
analysis should certainly not be enough to describe compactifications with
radii much smaller than the 11 dimensional Planck length, as we have 
attempted to do here. It is not clear if
anything can be done beyond this level without knowing a quantum
formulation for M theory. Perhaps a systematic analysis of the membrane
corrections to type IIA string theory at non-zero coupling, for instance,
may be possible just by assuming that M theory is supermembrane theory. 
Even at the low-energy level, the description of some string theory
effects, such as enhanced $U(n)$ gauge symmetry, in M theory is still
not clear. These effects may arise only at the quantum level, but they may
also correspond to classical effects which have not yet been investigated.
In any case, we hope that our work will lead to a better understanding of M
theory and of the correct way to formulate it.

We end with an amusing numerical 
observation about $p$-branes in M theory.
In our discussion we used the fact that membranes
in M theory can have boundaries on fivebranes, or on other membranes.
Let us raise the question 
whether membranes can have boundaries on any other, so far
unknown, $p$-branes in M theory. The boundary of a membrane is a string, so
it should couple to a 2-form $B_{\alpha \beta}$ in the worldvolume theory
of the $p$-brane. A BPS saturated $p$-brane breaks half the
supersymmetries. In 11 (or 10) dimensions this means it should have 16
normalizable fermionic zero modes, and correspondingly (by supersymmetry)
8 normalizable bosonic zero modes. These zero
modes correspond to the bosonic physical degrees of freedom on the 
worldvolume of the $p$-brane. If we assume that the only degrees of freedom
on the $p$-brane in 11 dimensions are the 2-form and the transverse 
excitations (corresponding to the motion of the $p$-brane in 11
dimensions), we find the equation
\eqn\degs{{{((p+1)-2)((p+1)-3)}\over 2} + (11 - (p+1)) = 8,}
whose solutions give $p=2$ or $p=3$. If $p=5$ there is an additional
possibility of having a self-dual or anti-self-dual 2-form. This halves the
number of degrees of freedom in the 2-form and leads to the additional
solution $p=5$. Allowing additional worldvolume gauge fields does not add 
any new solutions. 
The solutions $p=2,5$ correspond to the known $p$-branes of
M theory, but the interpretation of the $p=3$ solution is not clear. Of
course, such ``hand-waving'' arguments do not prove that such 3-branes
indeed exist (there are no corresponding solitons of 11 dimensional
supergravity as far as we know). 
For instance, using a similar computation for strings, 
whose ends couple to 1-form
fields, we would find that Dirichlet $p$-branes with any value of $p$ 
are possible. However, only half of the Dirichlet
$p$-branes of type II theories, and less in type I theory, can
consistently couple to superstrings. If such a 3-brane
exists, it would look like a 12 dimensional object, since a 2-form field
in $3+1$ dimensions is dual to a scalar. As in our analysis of strings
and 2-branes in section 2, the membrane could then be interpreted as a
3-brane wrapped around the twelfth dimension, and the interaction of the
membrane with the 3-brane would be simply
a Nambu-Goto type interaction for the 3-brane.
The relation of this observation
to recent discussions of 12 dimensional theories \refs{\vafa,\km,\tseytlin}
is still not clear.

\centerline{ }
\centerline{\bf Acknowledgments}

We thank O. Ganor, Y. Oz, J. Pawelczyk and P. K. Townsend for
discussions, and M.~B. Green for
comments and for reading this manuscript.

\listrefs

\end